\documentclass[reprint, amsmath,amssymb, aps, prb]{revtex4-1}

\usepackage{graphicx}% Include figure files
\usepackage{dcolumn}% Align table columns on decimal point
\usepackage{bm}% bold math
\usepackage{enumitem, changes}
\usepackage[hidelinks]{hyperref}% add hypertext capabilities
\usepackage{caption}
\usepackage{subcaption}

\DeclareUnicodeCharacter{202F}{!!!!FIX ME!!!!}

\begin{document}

% \title{Stress enabled prismatic dislocation loop nucleation in aluminum}
\title{Ab initio study of strain-driven vacancy clustering in aluminum}

\author{Sayan Bhowmik}
\affiliation{College of Engineering, Georgia Institute of Technology, Atlanta, Georgia 30332, USA}
\author{Andrew J. Medford}
\affiliation{College of Engineering, Georgia Institute of Technology, Atlanta, Georgia 30332, USA}
\author{Phanish Suryanarayana}
\email{phanish.suryanarayana@ce.gatech.edu}
\affiliation{College of Engineering, Georgia Institute of Technology, Atlanta, Georgia 30332, USA}
\author{Abhiraj Sharma}
\affiliation{Physics Division, Lawrence Livermore National Laboratory, Livermore, California 94550, USA}
\author{John E. Pask}
\affiliation{Physics Division, Lawrence Livermore National Laboratory, Livermore, California 94550, USA}

\begin{abstract}
We present a first principles investigation of strain-driven vacancy clustering in aluminum. Specifically,  we perform Kohn-Sham density functional theory calculations to study the influence of hydrostatic strains on clustering in tri-, quad-, and heptavacancies. We find that compressive strains are a key driving force for vacancy aggregation, particularly for collapse of clusters  on the (111) plane, consistent with prior experimental observations of vacancy clusters on this plane. Notably, we find that the heptavacancy on the  (111) plane collapses to form a prismatic dislocation loop for hydrostatic compressive strains exceeding 5\%, highlighting the critical role of such strains in prismatic dislocation loop nucleation in aluminum.
\end{abstract}

\maketitle

%%%%%%%%%%%%%%%%%%%%%%%%%%%%%%%%%%%%%%%%%%%%%%%%%%%%%%%%%%%%%%%%%%%%%%%%
%%%%%%%%%%%%%%%%%%%%%%%%%%%%%%%%%%%%%%%%%%%%%%%%%%%%%%%%%%%%%%%%%%%%%%%%
%%%%%%%%%%%%%%%%%%%%%%%%%%%%%%%%%%%%%%%%%%%%%%%%%%%%%%%%%%%%%%%%%%%%%%%%

\section{Introduction \label{Sec:Introduction}}

A vacancy is a point defect in a crystal lattice where an atom is missing from the site that it would normally occupy \cite{hull}. Once a critical concentration threshold is exceeded, vacancies can migrate and coalesce into clusters, typically during non-equilibrium processing or under extreme conditions such as rapid quenching, irradiation, or severe plastic deformation \cite{hull}.  After aggregation, these clusters can transform into more stable defect structures; in face-centered cubic (FCC) aluminum, for example, vacancy clusters may collapse into stacking-fault tetrahedra, voids, or prismatic dislocation loops \cite{hull}. The latter are extended defects typically observed on the (111) plane  \cite{PureAlExpt, AlMgAlloyExpt}, where they impede dislocation motion, contribute to hardening, and serve as key indicators of microstructural evolution.   However, the fundamental mechanisms governing vacancy clustering in aluminum, including the collapse of clusters into prismatic dislocation loops, are not well established. 

Kohn-Sham density functional theory (DFT) \cite{InhomElecGas, KS_DFT} is a widely used ab initio method for understanding and predicting material behavior.  It has been employed to study the energetics of point defects in aluminum, including monovacancies \cite{CarlingAlVac, Carling2, UesugiAlVac, Vac5prb, VacCluster8DFT, CONNETABLE2020869}, divacancies \cite{CarlingAlVac, Carling2, UesugiAlVac, Vac5prb, VacCluster8DFT}, and, more recently, clusters containing up to eight vacancies \cite{Vac5prb, VacCluster8DFT}.  It has been found that the tetrahedral pentavacancy is energetically favorable relative to monovacancies, binding and relaxing into a stacking-fault tetrahedron \cite{Vac5prb, VacCluster8DFT}. In addition, it has been found that platelet-type hexa-, hepta-, and octavacancies, which are (111) plane clusters capable of collapsing into prismatic loops for the heptavacancy and larger, are energetically unfavorable compared to monovacancies \cite{VacCluster8DFT}. While these ab initio studies have provided valuable insights into vacancy clustering in aluminum, they have not addressed the effect of strains, which are commonly encountered in the aforementioned environments where such clusters can form and evolve. Furthermore, the reason why the (111) plane is favorable for vacancy clustering, particularly for the formation of dislocation loops as observed in experiments \cite{PureAlExpt, AlMgAlloyExpt}, remains unclear. In addition, the minimum size of a vacancy cluster capable of forming a prismatic loop has not been determined. Also, there are inconsistencies among the reported results. For example, while some studies report that divacancies are energetically unfavorable relative to monovacancies \cite{Vac5prb, VacCluster8DFT}, others have found them to be favorable \cite{CarlingAlVac, Carling2, UesugiAlVac}. A similar disagreement exists regarding the relative energetics of the tetrahedral quadvacancy \cite{Vac5prb, VacCluster8DFT}.

Orbital-free DFT is a simplified version of Kohn-Sham DFT wherein the orbital-dependent kinetic energy functional is approximated by a functional of the electron density \cite{parr1989density, wang2000orbital}. The large computational cost associated with Kohn-Sham DFT calculations has motivated the use of orbital-free DFT for the study of vacancy clustering in aluminum \cite{QCOFDFT, VikramQuadPRB, CarterTriPCCP, VikramPRL_Strain, QCOFDFT_WEGC, VikramQuadWEGC}. It has been found that hydrostatic strains make di- \cite{CarterTriPCCP, VikramPRL_Strain, Vikram_PRSCA} and trivacancies \cite{CarterTriPCCP} more energetically favorable relative to monovacancies. In addition, quadvacancies on the (111) plane, i.e., platelet quadvacancies, are found to be the most favorable in terms of binding, in contrast to Kohn-Sham DFT results, which indicate that they are not energetically favorable relative to monovacancies \cite{Vac5prb, VacCluster8DFT}. It has also been found that, depending on the choice of the kinetic energy functional in orbital-free DFT, the heptavacancy \cite{VikramQuadPRB} or a nineteen vacancy cluster \cite{VikramQuadWEGC} on the (111) plane collapses to form a prismatic dislocation loop. Though these studies have offered important insights into vacancy clustering in aluminum, the accuracy of orbital-free DFT is limited in practice, primarily due to the use of approximate kinetic energy functionals and local pseudopotentials largely designed for equilibrium crystal configurations. Moreover, the effect of hydrostatic  strains on the formation of prismatic dislocation loops has not been investigated. Also, the origin of the (111) plane's favorability for clustering has also not been clearly established. The limitations of orbital-free DFT studies, together with those of previous Kohn-Sham DFT studies, motivate the present effort.

In the present work, we conduct a first-principles study of vacancy clustering in aluminum using Kohn-Sham DFT, examining the effect of hydrostatic strains on tri-, quad-, and heptavacancies. We find that compressive strains drive vacancy aggregation, particularly on the (111) plane where collapse occurs. Notably, the heptavacancy on this plane collapses into a prismatic dislocation loop under hydrostatic compressive strains exceeding 5\%, underscoring the critical role of such strains in prismatic dislocation loop nucleation in aluminum.

%%%%%%%%%%%%%%%%%%%%%%%%%%%%%%%%%%%%%%%%%%%%%%%%%%%%%%%%%%%%%%%%%%%%%%%%
%%%%%%%%%%%%%%%%%%%%%%%%%%%%%%%%%%%%%%%%%%%%%%%%%%%%%%%%%%%%%%%%%%%%%%%%
%%%%%%%%%%%%%%%%%%%%%%%%%%%%%%%%%%%%%%%%%%%%%%%%%%%%%%%%%%%%%%%%%%%%%%%%

\section{Systems and methods\label{Sec:SysMeth}}
Consider the partitioning  of the $M \geq 2$ vacancy cluster into $K \geq 2$ smaller vacancy clusters: $M = \sum_{k=1}^K M_k$, where $M_k \in \{1, \ldots, M-1 \}$. The vacancy binding enthalpy, an experimentally relevant measure of vacancy clustering, is defined as:
\begin{equation} \label{Eq:Hvb}
    H_{vb}(M,M_k) = \sum_{k=1}^K H_{vf}(M_k, P)  - H_{vf}(M, P) \,, 
\end{equation}  
where $P$ is the pressure, and the vacancy formation enthalpy, an experimentally relevant measure of vacancy formation in the bulk,  is defined as:
\begin{align} 
    H_{vf}(\hat{M}, P) = E_{vf}(\hat{M}, P) + P \,  V_{vf}(\hat{M}, P) \,. \label{Eq:Hvf}
\end{align}
Above, the vacancy formation energy at constant pressure is defined as:
\begin{align}
E_{vf}(\hat{M}, P)  = & E \left(N-\hat{M},\hat{M},P \right) \nonumber \\ 
& - \left( \frac{N-\hat{M}}{N}\right) E(N,0,P) \,, 
\end{align}
where $E(N-\hat{M},\hat{M},P)$ denotes the ground-state energy of the system consisting of $N$ lattice sites, with $(N-\hat{M})$ occupied and $\hat{M}$ vacant, evaluated at the pressure $P$. The change in volume due to the introduction of the vacancy in the bulk can be written as:
\begin{align}
V_{vf}(\hat{M},P)  = &  V(N-\hat{M},\hat{M},P) \nonumber \\  
& - \frac{N-\hat{M}}{N}V(N,0,P) \,,
\end{align} 
where $V(N-\hat{M},\hat{M},P)$ denotes the volume of a system consisting of $N$ lattice sites, with $(N-\hat{M})$ occupied and $\hat{M}$ vacant, evaluated at the pressure $P$.

Denoting $V(N, 0, P)$ as $V$, with $V$ and $P$ related through the bulk equation of state, it can be shown that to first order the vacancy formation energy at constant pressure  \cite{DeVita_1991}
\begin{align}
     E_{vf}(\hat{M},P) = \widetilde{E}_{vf}(\hat{M},V) - P V_{vf}({\hat{M}},P) \,, \label{eq_EvfP_exp}
\end{align}
where the vacancy formation energy 
\begin{align}  
    \widetilde{E}_{vf}(M, V) = & E \left(N-M,M,\frac{N-M}{N} V \right) \nonumber \\
    &- \left( \frac{N-M}{N}\right)E(N,0,V) \,. \label{Eq:EvfV}
\end{align} 
It follows from Eqs.~\ref{Eq:Hvf} and~\ref{eq_EvfP_exp} that, to first order: 
\begin{align} 
H_{vf}(\hat{M}, P) = \widetilde{E}_{vf}(\hat{M},V) \,, \label{Eq:EtvfHvf}
\end{align}  
so that $H_{vf}$ can be well approximated by $\widetilde{E}_{vf}$ when higher-order contributions can be neglected. The vacancy formation energy at constant volume can itself can be expanded to first order as:
\begin{align}
    \widetilde{E}_{vf}(\hat{M},V) = E_{vf}(\hat{M},V) + P \frac{\hat{M}}{N} V(N,0,P) \,, \label{EvfEvft}
\end{align}
where 
\begin{align}
E_{vf}(\hat{M}, V) =&  E(N-\hat{M},\hat{M},V)  \nonumber \\
 & - \left( \frac{N-\hat{M}}{N}\right) E(N,0,V) \,,
\end{align}
with $E(N-\hat{M},\hat{M},V)$ being  the ground state energy of the system consisting of $N$ lattice sites, with  $(N-\hat{M})$ occupied and $\hat{M}$ vacant, evaluated at the volume $V$. It therefore follows from Eqs.~\ref{Eq:Hvb}, \ref{Eq:EtvfHvf}, and \ref{EvfEvft} that, to first order:
\begin{align}
H_{vb}(M,M_k)  = E_{vb}(M,M_k) \,, \label{HvbEvb}
\end{align}
where the vacancy binding energy 
\begin{equation} \label{Eq:BE}
    E_{vb}(M,M_k) = \sum_{k=1}^K E_{vf}(M_k, V)  - E_{vf}(M, V) \,.
\end{equation}
In arriving at Eq.~\ref{HvbEvb}, we use the relation $M = \sum_{k=1}^K M_k$, which causes the pressure term in Eq.~\ref{EvfEvft} to vanish upon substitution into Eq.~\ref{Eq:Hvb}. The detailed derivation, including higher-order terms, can be found in the Supplementary Material \cite{Supplemental}. In summary,  $H_{vb}$ can be well approximated by $E_{vb}$ when higher-order contributions can be neglected.

A positive binding enthalpy/energy indicates that vacancy aggregation is energetically favorable, while a negative value implies the opposite. Moreover, larger positive values correspond to a greater tendency for vacancies to aggregate. The above definitions of binding enthalpy/energy allow their evaluation not only with monovacancies as the reference, as is commonly done, but also using various combinations of vacancy configurations as references. For instance, a quadvacancy with \( M = 4 \) can be partitioned into four monovacancies (\( K = 4 \) with \( M_1 = M_2 = M_3 = M_4 = 1 \)), two divacancies (\( K = 2 \) with \( M_1 = M_2 = 2 \)), or a trivacancy and a monovacancy (\( K = 2 \) with \( M_1 = 3 \), \( M_2 = 1 \)), each of which can serve as a reference for the binding energy/enthalpy. The motivation for calculating the binding energy/enthalpy using various reference vacancy configurations is to quantitatively assess the likelihood of a given cluster forming from those configurations, as multiple formation pathways may exist. In particular, borrowing ideas from reaction kinetics \cite{laidler_book}, variations in binding energies across reference states are expected to reflect differences in kinetic accessibility. Indeed, the transition state scaling relations used to estimate the barriers depends on the binding energies, yielding distinct barriers for each pathway, thereby influencing the corresponding rates. Moreover, for larger clusters, stepwise aggregation of smaller defects is expected to be more probable than simultaneous clustering. Multiple references thus better capture both thermodynamic driving forces and realistic kinetic pathways.

In this work, we consider the vacancy configurations listed in Table~\ref{VacPos}. In particular, we consider: (i) a monovacancy; (ii) two divacancy configurations, labeled \texttt{N} and \texttt{NN}, corresponding to vacancies in nearest-neighbor and next-nearest-neighbor positions, respectively; (iii) eight trivacancy configurations, labeled \texttt{A} through \texttt{H}, formed by adding a third vacancy to the \texttt{N} or \texttt{NN} divacancies in either a nearest-neighbor or next-nearest-neighbor position; (iv) fourteen quad-vacancy configurations, labeled \texttt{A1}, \texttt{G1} through \texttt{G9}, and \texttt{H1} through \texttt{H4}, formed by adding a fourth vacancy to the \texttt{A}, \texttt{G}, and \texttt{H} trivacancies, respectively, in the nearest-neighbor position; and (v) a heptavacancy configuration, labeled \texttt{A11}, formed by adding three vacancies to the \texttt{A1} quadvacancy in nearest-neighbor positions on the (111) plane, such that there are no atoms enclosed within the vacancy-defined boundary. While there are, in principle, infinitely many possible vacancy configurations, the selection in this work is guided by the interest in studying vacancy clustering, for which nearest- and next-nearest-neighbor arrangements are most relevant. Since this leads to a large number of quadvacancy configurations when starting from the chosen trivacancies, we focus on those formed from the \texttt{G} and \texttt{H} trivacancies due to their favorable energetics. We also include the \texttt{A1} quadvacancy in the analysis, as it lies on the experimentally observed (111) plane, with similar motivation for the inclusion of the \texttt{A11} heptavacancy. Note that penta- and hexa-vacancies are not included in the detailed analysis and are used only to verify cluster collapse on the (111) plane, since the heptavacancy is the smallest cluster capable of forming a prismatic dislocation loop, which is the focus of this work.

\begin{table}[htbp]%The best place to locate the table environment is directly after its first reference in text
\caption{\label{VacPos}%
Vacancy configurations considered, where $a$ denotes the lattice constant. The trivacancy \texttt{A}, quadvacancy \texttt{A1}, and heptavacancy \texttt{A11} all lie on the (111) plane.}
\begin{ruledtabular}
\begin{tabular}{lllll}
\textrm{Type}&
\multicolumn{4}{c}{\textrm{Position}}
\\
\colrule
\colrule
\textbf{mono} & (0,0,0) & \multicolumn{3}{l}{}\\

\colrule
\textbf{di} & \multicolumn{4}{l}{}\\
\colrule
\texttt{N} & (0,0,0) & ($a$/2,$a$/2,0) & \multicolumn{2}{l}{} \\
\texttt{NN} & (0,0,0) & ($a$,0,0) & \multicolumn{2}{l}{} \\
% NNN & (0,0,0) & (a/2,a,a/2) & \multicolumn{2}{l}{} \\

\colrule
\textbf{tri} & \multicolumn{4}{l}{}\\
\colrule
\texttt{A} & (0,0,0) & ($a$/2,$a$/2,0) & ($a$/2,0,$a$/2) & \\
\texttt{B} & ($a$/2,$a$/2,0) & ($a$,0,0) & (3$a$/2,0,$a$/2) & \\
\texttt{C} & (0,0,0) & ($a$/2,0,$a$/2) & ($a$,0,0) & \\
\texttt{D} & (0,0,0) & ($a$/2,0,$a$/2) & ($a$,0,$a$) & \\
\texttt{E} & (0,0,0) & ($a$,0,0) & (3$a$/2,0,$a$/2) & \\
\texttt{F} & (0,0,0) & ($a$,0,0) & ($a$,$a$/2,$a$/2) & \\
\texttt{G} & (0,0,0) & ($a$,0,0) & ($a$,0,$a$) & \\
\texttt{H} & (0,0,0) & ($a$,0,0) & (2$a$,0,0) & \\

\colrule
\textbf{quad} & \multicolumn{4}{l}{}\\
\colrule
\texttt{A1} & (0,0,0)& (0,$a$/2,$a$/2)& ($a$/2,$a$/2,0)& ($a$/2,$a$,$a$/2) \\
\texttt{G1} & (0,0,0) & ($a$,0,0) & ($a$,0,$a$) & (-$a$/2,$a$/2,0) \\
\texttt{G2} & (0,0,0) & ($a$,0,0) & ($a$,0,$a$) & (0,-$a$/2,$a$/2) \\
\texttt{G3} & (0,0,0) & ($a$,0,0) & ($a$,0,$a$) & (-$a$/2,0,$a$/2) \\
\texttt{G4} & (0,0,0) & ($a$,0,0) & ($a$,0,$a$) & (0,-$a$/2,-$a$/2) \\
\texttt{G5} & (0,0,0) & ($a$,0,0) & ($a$,0,$a$) & ($a$/2,-$a$/2,0) \\
\texttt{G6} & (0,0,0) & ($a$,0,0) & ($a$,0,$a$) & (-$a$/2,0,-$a$/2) \\
\texttt{G7} & (0,0,0) & ($a$,0,0) & ($a$,0,$a$) & ($a$/2,0,$a$/2) \\
\texttt{G8} & (0,0,0) & ($a$,0,0) & ($a$,0,$a$) & ($a$,-$a$/2,$a$/2) \\
\texttt{G9} & (0,0,0) & ($a$,0,0) & ($a$,0,$a$) & ($a$/2,0,-$a$/2) \\
\texttt{H1} & (0,0,0) & ($a$,0,0) & (2$a$,0,0) & (-$a$/2,-$a$/2,0) \\
\texttt{H2} & (0,0,0) & ($a$,0,0) & (2$a$,0,0) & (0,-$a$/2,$a$/2) \\
\texttt{H3} & (0,0,0) & ($a$,0,0) & (2$a$,0,0) & ($a$/2,-$a$/2,0) \\
\texttt{H4} & (0,0,0) & ($a$,0,0) & (2$a$,0,0) & ($a$,-$a$/2,$a$/2) \\

\colrule
\textbf{hepta} & \multicolumn{4}{l}{}\\
% \textbf{(expt)} & \multicolumn{4}{l}{}\\
\colrule
\texttt{A11} & (0,0,$a$) & ($a$/2,0,$a$/2) & ($a$,0,0) & (0,$a$/2,$a$/2) \\
& $a$/2,$a$/2,0) & ($a$/2,-$a$/2,$a$) & ($a$,-$a$/2, $a$/2) & \\

\end{tabular}
\end{ruledtabular}
\end{table}

We perform Kohn-Sham DFT calculations  using the graphics processing unit (GPU)-accelerated version \cite{SPARC_GPU, 10.1063/5.0260892} of the SPARC electronic structure code \cite{SPARC_SoftX, zhang2024sparc}. SPARC is a real-space code that employs high-order finite-difference discretization~\cite{SPARC_1, SPARC_2}, with convergence to the infinite-basis limit governed by a single parameter, the grid spacing.   In all calculations, we employ the ONCV pseudopotential \cite{ONCV} with nonlinear core corrections from the SPMS table \cite{SPMS}, which has 3 electrons in valence. Given its small $\Delta$-factor~\cite{delta_Test_Sci, SPMS}, which quantifies the error introduced by the pseudopotential in the equation of state for hydrostatic strains,  the chosen pseudopotential is expected to be accurate for the conditions considered. The computed equation of state is provided in the Supplementary Material \cite{Supplemental}. In certain instances, we also consider the stringent variant of the  \texttt{PseudoDOJO} ONCV pseudopotential \cite{dojo}, which is significantly harder than the chosen SPMS pseudopotential and so provides a strong, independent check. We employ the Perdew-Burke-Ernzerhof (PBE) \cite{GGA_PBE} exchange-correlation functional, which is based on the generalized gradient approximation (GGA).     In select cases, we employ the local density approximation (LDA) \cite{KS_DFT} to examine the effects of the exchange-correlation functional, using the corresponding LDA version of the SPMS pseudopotential. In this setting, i.e., PBE exchange-correlation and SPMS pseudopotential, we determine the equilibrium lattice constant for FCC aluminum to be  $a=7.629$ bohr. The mono-, di-, tri-, and quadvacancies  are introduced in 864-atom cells, formed by $6 \times 6 \times 6$ FCC unit cells, while the heptavacancy is introduced in a 2048-atom cell, formed by $8 \times 8 \times 8$ FCC unit cells. The finite cell size effects are discussed in the Supplementary Material \cite{Supplemental}. We employ a real-space grid spacing of 0.35 bohr, with $4 \times 4 \times 4$ and $3 \times 3 \times 3$ Monkhorst-Pack \cite{MonkhorstPack} grids for Brillouin zone integration in the case of the 864-atom and 2048-atom cells, respectively. In all cases, structural ionic relaxation is performed until the forces on all the atoms fall below $10^{-4}$ ha/bohr. For the enthalpy calculations, isotropic volume relaxations are additionally performed until the pressure converges to within $0.01$~GPa of the target value. The chosen system sizes and parameters ensure convergence of total energies and forces to within $2 \times 10^{-4}$ ha/atom and $2 \times 10^{-4}$ ha/bohr, respectively, which translates  to convergence of vacancy binding energies ($E_{vb}$) to within $5$ meV. Indeed, the relatively high accuracy of $E_{vb}$ arises from significant error cancellation, up to multiple orders of magnitude, since the same simulation cell and discretization grid are employed throughout in its evaluation (Eq.~\ref{Eq:BE}).

%%%%%%%%%%%%%%%%%%%%%%%%%%%%%%%%%%%%%%%%%%%%%%%%%%%%%%%%%%%%%%%%%%%%%%%%
%%%%%%%%%%%%%%%%%%%%%%%%%%%%%%%%%%%%%%%%%%%%%%%%%%%%%%%%%%%%%%%%%%%%%%%%
%%%%%%%%%%%%%%%%%%%%%%%%%%%%%%%%%%%%%%%%%%%%%%%%%%%%%%%%%%%%%%%%%%%%%%%%

\section{Results and discussion\label{Sec:Results}}

\begin{figure}[htbp]
    \centering
    \includegraphics[width=0.40\textwidth]{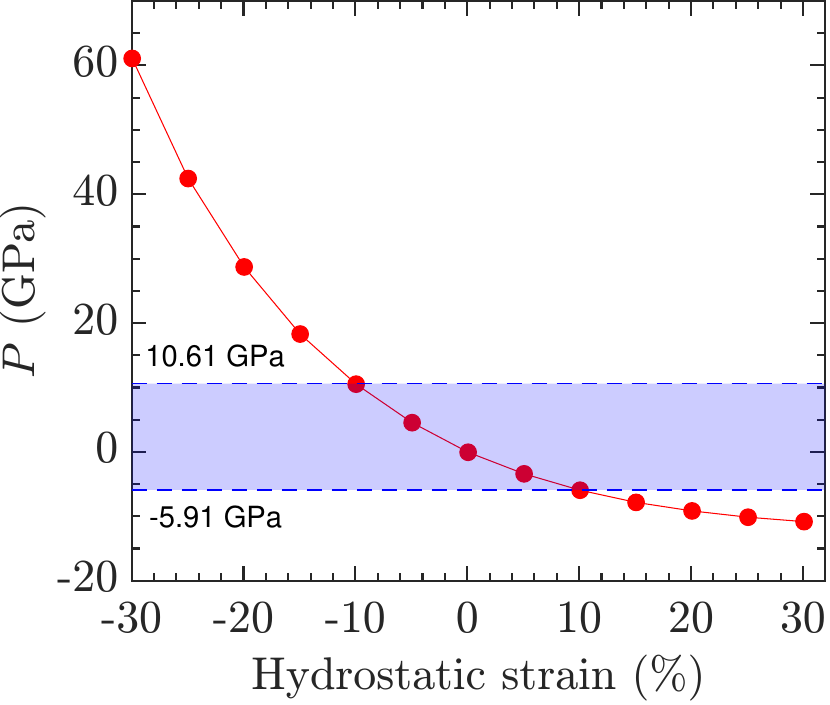}
    \caption{Equation of state for bulk aluminum, with the shaded area indicating the range of pressures studied.}
    \label{fig:EOS}
\end{figure}

We now use the framework described in the previous section to study the effect of isotropic hydrostatic strains on vacancy clustering in  aluminum. In particular, we perform Kohn-Sham DFT calculations for the unstrained system, as well as for 5\% and 10\% compressive and tensile hydrostatic strains.The hydrostatic strain is introduced by applying the following deformation gradient to the lattice vectors:
\begin{align}
\mathbf{F} = 
\begin{bmatrix}
\lambda & 0 & 0 \\
0 & \lambda & 0 \\
0 & 0 & \lambda
\end{bmatrix} \,, \, \lambda = \sqrt[3]{1+ \frac{\text{hydrostatic strain (\%)}}{100}} \,,
\end{align} 
the procedure for which is available in the literature \cite{strain_engg_graphene}. The maximum bulk pressure so encountered is $\sim 10.6$ GPa, as determined from the equation of state of bulk aluminum (Fig.~\ref{fig:EOS}), which is well within the range commonly encountered in spallation, shock, and laser ablation experiments~\cite{spallExpt, laserPlasmaExpt}. The choice of  hydrostatic strains is motivated by their comparatively greater impact on vacancy properties in aluminum \cite{ghosh2019electronic, VikramPRL_Strain, Vikram_PRSCA, VikramPollockDeformation, CONNETABLE2020869}. The vacancy binding energies  are computed for all the vacancy configurations, whereas the binding enthalpies are computed for select cases of interest:  \texttt{NN} divacancy, \texttt{A} trivacancy, and \texttt{A1} quadvacancy. The results presented here focus on the unstrained and compressive strain cases.  The detailed data, including results for tensile strains, are provided in the Supplementary Material \cite{Supplemental}. The results for the vacancy formation energy at constant volume and formation enthalpy, quantities of interest for vacancies though not directly related to vacancy clustering, are provided in Appendix~\ref{App:VFE}.

%%%%%%%%%%%%%%
%%%%%%%%%%%%%%%%%

%%%%%%%%%%%%%%%%
\begin{figure}[htbp]
\includegraphics[width=0.44\textwidth, keepaspectratio]{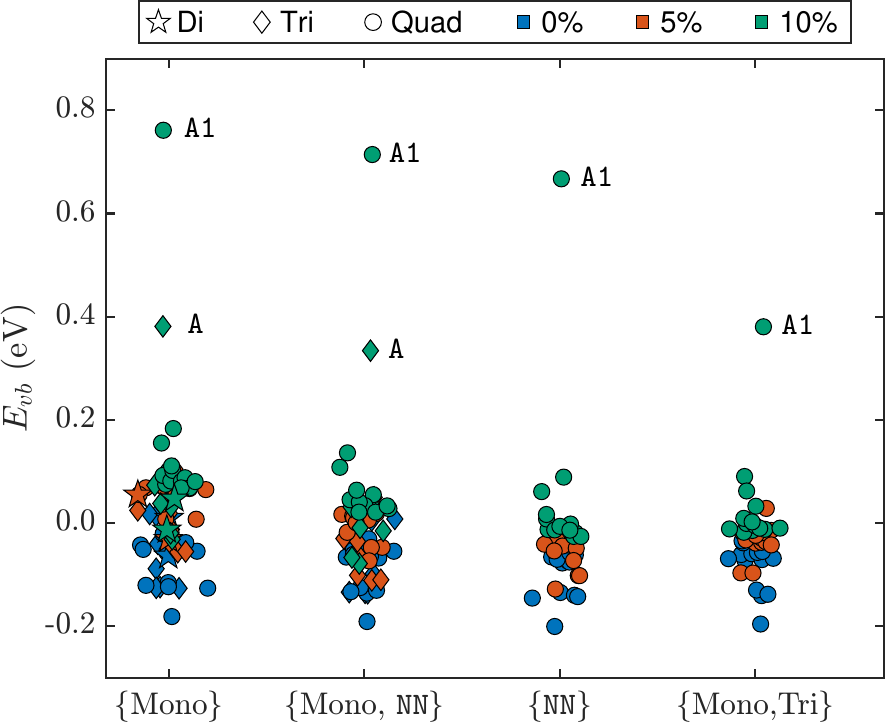}
\caption{Binding energy for the various vacancy configurations in the unstrained, 5\% compressive, and 10\% compressive strain cases. The labels along the x-axis denote the  vacancy clusters that are used as reference. For instance, \texttt{\{Mono,NN\}} indicates that the binding energy is computed with respect to a combination of a monovacancy and an \texttt{NN} divacancy in the case of a trivacancy, whereas for a quadvacancy, it refers to a combination of two monovacancies and an \texttt{NN} divacancy. \label{Fig:VBE}}
\end{figure}
%%%%%%%%%%%%%%%%

In Fig.~\ref{Fig:VBE}, we present the binding energy for the di-, tri-, and quadvacancies. The divacancy binding energy is computed using two monovacancies as reference; the trivacancy binding energy using three monovacancies or a combination of a monovacancy and the \texttt{NN} divacancy as reference; and the quadvacancy binding energy using four monovacancies, a combination of the trivacancy from which it was derived and a monovacancy, or two \texttt{NN} divacancies as reference.  We find that the unstrained divacancy binding energy is positive for the \texttt{NN} divacancy and negative for the \texttt{N} divacancy.  Indeed, the binding energy for the \texttt{N} divacancy remains negative even on the application of strains, hence it is omitted from the analysis for tri- and quadvacancies. Among the trivacancies, only \texttt{G} and \texttt{H} have positive binding energies in the unstrained configuration, which is consistent with previous orbital-free DFT calculations \cite{CarterTriPCCP}. None of the quadvacancies have positive binding energies in the unstrained configuration. In all instances, we observe that presence of compressive strains makes the aggregation of vacancies more favorable, which is also the case for tensile strains, albeit the effects are not as substantial (Supplementary Material \cite{Supplemental}). These trends for the binding energy variation with hydrostatic strains are in agreement with orbital-free DFT results for mono-, di-, and tri-vacancies \cite{CarterTriPCCP, VikramPRL_Strain}.  We also observe that the \texttt{A} trivacancy and \texttt{A1} quadvacancy are outliers for 10\% compressive strain, exhibiting significantly higher binding energies compared to the other tri- and quadvacancies, respectively.  We have verified that this remains the case upon changing the exchange-correlation functional to LDA or upon using the  \texttt{PseudoDOJO} pseudopotential. Note that the \texttt{A} trivacancy and \texttt{A1} quadvacancy are also outliers in the formation energy at 10\% compressive strain (Appendix~\ref{App:VFE}). Indeed, the binding energy of the \texttt{A1} quadvacancy relative to the \texttt{A} trivacancy and a monovacancy at 10\% compressive strain would have been even larger were it not for the relatively low formation energy of the \texttt{A} trivacancy.

The vacancy binding enthalpy is in very good agreement with the binding energy (Supplementary Material \cite{Supplemental}), indicating that the higher order effects do not play a significant role in the binding enthalpy for the cases considered.  As a result, the sign and variation of the binding enthalpy with strain follows the same trend as the binding energy. Indeed, the \texttt{A} trivacancy and \texttt{A1} quadvacancy  are also outliers in the binding enthalpy values for 10\% compressive strain, further confirming the physical significance of the results.

The  divacancy binding energies of $-0.064$ and $0.009$ eV for the unstrained \texttt{N} and \texttt{NN} divacancies, respectively, are in good agreement with previous DFT calculations, which report values $-0.1$ to $-0.05$ eV for \texttt{N} and $0.00$ eV for \texttt{NN}.\cite{Vac5prb,VacCluster8DFT}. Indeed, the binding energies  for \texttt{N} and \texttt{NN}  remain negative and positive, respectively, on the application of strains.  The binding enthalpy of $0.005$ eV for the unstrained  \texttt{NN} divacancy is also in good agreement with previous DFT calculations, which report values from $0.004$ to $0.04$ eV\cite{CarlingAlVac, UesugiAlVac,Kaxiras2}. Again, the binding enthalpy remains positive on the application of strains. The experimental value ranges from $0.17$ to $0.30$~eV~\cite{Vac_Conc_Fluss, ExptDiVacBinding, HEHENKAMP1994907}, which is substantially larger than the computed \texttt{NN} binding energy/enthalpy, but is qualitatively consistent in indicating that the aggregation of two isolated monovacancies into a divacancy is energetically preferred. The binding energies of the unstrained  \texttt{A} trivacancy and \texttt{A1} quadvacancy with monovacancies as reference, computed as $-0.089$ and $-0.124$ eV, respectively, are in reasonable agreement with previous DFT values of $-0.06$ and $-0.058$ eV \cite{VacCluster8DFT}. The differences between the values reported in previous Kohn-Sham calculations and those obtained in the present work can be attributed, in part, to finite-size effects in the earlier studies. The trivacancy binding energies in the absence of any strain are in reasonable agreement with previous orbital-free DFT calculations \cite{CarterTriPCCP}, with a maximum difference of $0.057$ eV for the \texttt{H} trivacancy. The binding energy for the \texttt{G7} quadvacancy is $-0.182$ eV, while the values predicted by orbital-free DFT, depending on the choice of kinetic energy functional,  are $0.662$ eV and $0.282$ eV \cite{VikramQuadPRB, VikramQuadWEGC}. The difference in the values are likely a consequence of the limitations of orbital-free DFT, namely approximate kinetic energy functionals and local pseudopotentials, the latter primarily developed for unstrained crystal configurations. In particular, the Kohn-Sham DFT calculations employed in this work use the exact form of the non-interacting electronic kinetic energy expressed in terms of single-particle wavefunctions, whereas orbital-free DFT relies on density-based approximations.

\begin{figure*}[htbp]
    \centering
    \includegraphics[width=0.99\textwidth, keepaspectratio]{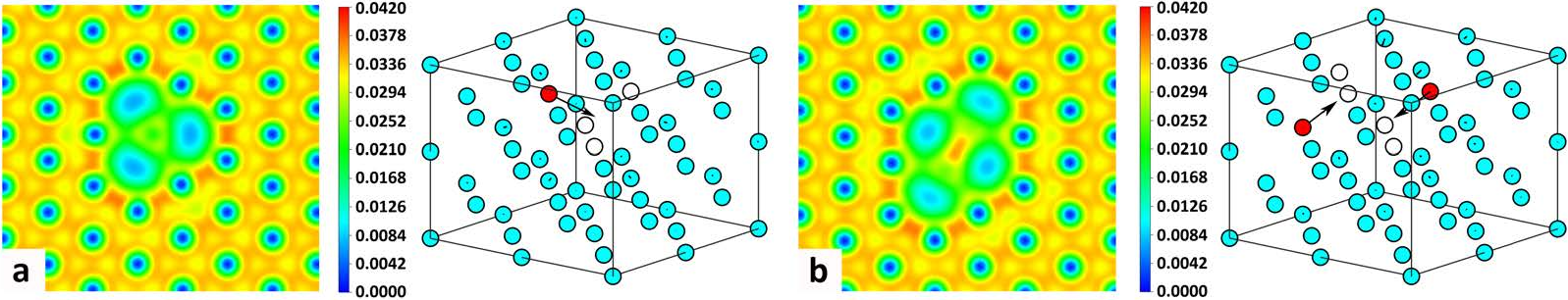}
    \caption{Contours of electron density on the $(111)$ plane and displacement of atoms for the  collapsed (a) \texttt{A} trivacancy and (b) \texttt{A1} quadvacancy clusters at 10\% compressive strain. White spheres indicate vacancy locations, blue spheres represent initial atomic positions, red spheres highlight atoms with the largest displacements, and arrows show the direction and magnitude of atomic movement. \label{Fig:TriQuad}}
\end{figure*}

The above results for the vacancy  binding energies and enthalpies of tri- and quadvacancies indicate that compressive strains are a key driving force for vacancy clustering, particularly on the (111) plane, which is the dominant clustering plane observed in experiments \cite{PureAlExpt, AlMgAlloyExpt}. This is especially true for the \texttt{A} trivacancy and \texttt{A1} quadvacancy, both located on the (111) plane, which exhibit substantially higher binding energies than the other tri- and quadvacancies, respectively, for 10\% compressive strain. To better understand the origin of these noticeably larger values, Fig.~\ref{Fig:TriQuad} shows the electron density contours on the (111) plane and the atomic displacements upon relaxation for the \texttt{A} trivacancy and \texttt{A1} quadvacancy under 10\% compressive strain. We observe significant displacements in one atom for the trivacancy and two atoms for the quadvacancy, indicating collapse of the vacancy cluster. Specifically, in the trivacancy, the highlighted atom undergoes a displacement of $3.18$ bohr, with the next largest displacement being $0.27$ bohr. In the quadvacancy, the two highlighted atoms exhibit displacements of $2.82$ bohr, while the next largest displacement is $0.569$ bohr. These large displacements lead to a significant increase in electron density within the vacancy cluster, which is otherwise absent, resulting in substantial lowering of the energy and therefore enhanced binding energies. The large atomic displacements can be associated with an instability, which is amenable to analysis using Kohn-Sham DFT.  This effect cannot be attributed to the (111) planes moving closer together, as compressive hydrostatic strain causes all planes to do so, with nothing unique about the (111) plane. Note that upon relaxing the cell to release the pressure, the atoms do not return to the unstrained positions, indicating a permanent collapse. Note also that the penta- and hexavacancies located on the (111) plane similarly collapse at 10\% compressive strain (Supplementary Material \cite{Supplemental}).

The propensity for vacancy clusters on the (111) plane to collapse under hydrostatic strains motivates the study of the \texttt{A11} heptavacancy, which is the smallest vacancy cluster on the  (111) plane that can collapse to form a prismatic dislocation loop. The binding energy for the heptavacancy, referenced to monovacancies, is computed to be $1.105$, $1.453$, and $2.026$~eV for the unstrained, 5\% compressive, and 10\% compressive strain cases, respectively. These results indicate that the clustering of seven isolated monovacancies to form a heptavacancy on the (111) plane  is energetically favorable, becoming more so under compressive hydrostatic strains, the trend being consistent with that observed for the other vacancy configurations. Although previous Kohn-Sham DFT results for the unstrained heptavacancy also yield a positive binding energy of $0.004$~eV \cite{VacCluster8DFT}, this value is considerably smaller than that obtained in the present work. The difference can be attributed, in large part, to finite-size effects in the earlier study, which used a $4 \times 4 \times 4$ FCC cell compared to the $8 \times 8 \times 8$ cell used here, the former being eight times smaller in size.

\begin{figure*}[htbp]
\includegraphics[width=0.7\textwidth, keepaspectratio]{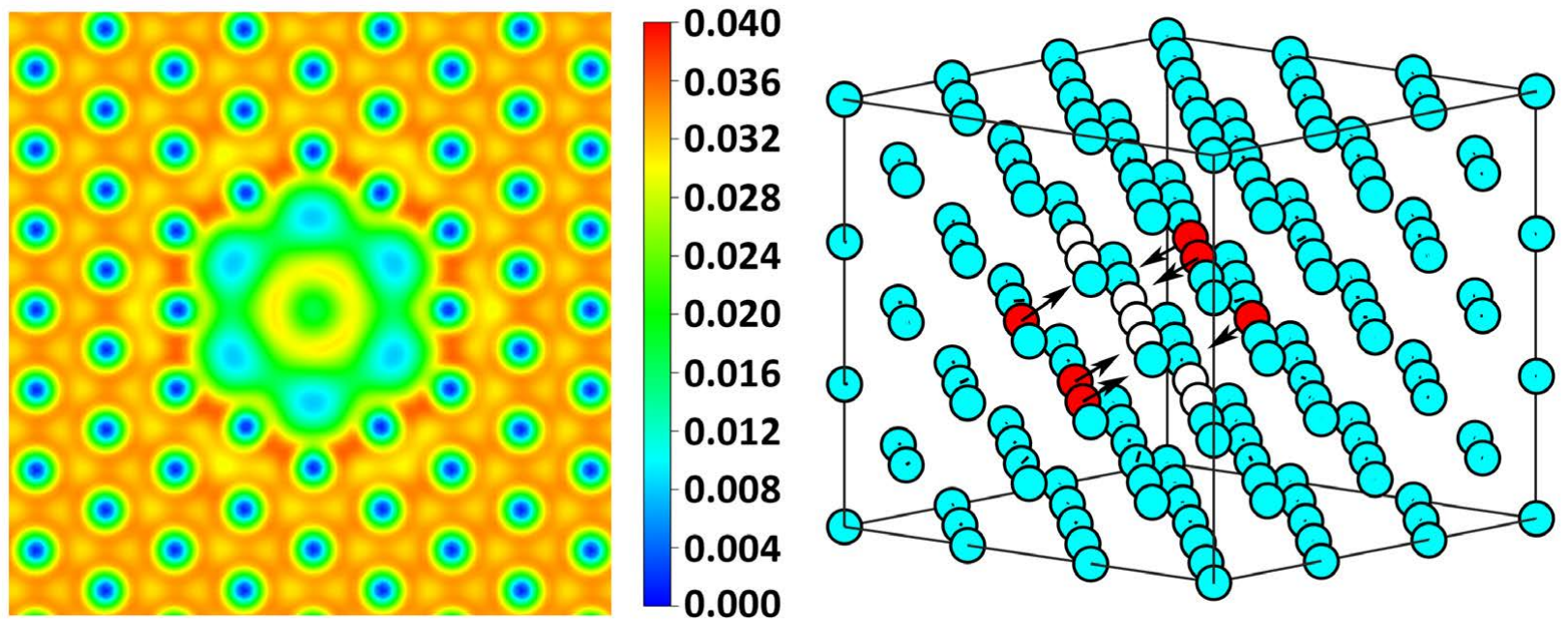}
\caption{Contours of electron density on the $(111)$ plane and displacement of atoms for the  collapsed heptavacancy cluster, i.e., prismatic dislocation loop, at 10\% compressive strain. White spheres indicate vacancy locations, blue spheres represent initial atomic positions, red spheres highlight atoms with the largest displacements, and arrows show the direction and magnitude of atomic movement. \label{Fig:Prismatic}}
\end{figure*}

In Fig.~\ref{Fig:Prismatic}, we show the electron density contours on the $(111)$ plane for the heptavacancy at 10\% compressive strain. In addition, we show the atomic displacements upon relaxation . We observe that similar to the \texttt{A} trivacancy and \texttt{A1} quadvacancy, certain atoms undergo displacements that are significantly larger than the others. In particular, six atoms undergo a large displacement of $2.085$~bohr, with the next largest displacement being $0.583$~bohr. The resulting atomic configuration corresponds to a prismatic dislocation loop, as confirmed by the atomic positions on the $(112)$ plane (Supplementary Material \cite{Supplemental}). While this loop is absent in the unstrained configuration, it also forms under 5\% compressive strain.  The trend of smaller compressive strains required for heptavacancy collapse, compared to the trivacancy and quadvacancy, suggests that the strain needed to initiate collapse decreases with increasing vacancy cluster size, indicating that a sufficiently large cluster may collapse even in the unstrained configuration. Similar to the \texttt{A} trivacancy and \texttt{A1} quadvacancy,  the large atomic displacements lead to significant increase in the electron density on the (111) plane. Indeed, experiments have found  that dislocation loops can trap electrons, which can significantly alter material properties \cite{elecTrap_disloc}.  Orbital-free DFT calculations in the literature have suggested that in the unstrained configuration, depending on the choice of kinetic energy functional, the heptavacancy and 19-vacancy cluster can collapse to form a prismatic dislocation loop \cite{VikramQuadPRB, VikramQuadWEGC}. The more accurate Kohn-Sham DFT results presented here indicate that the heptavacancy can form a prismatic dislocation loop, though only in the presence of sufficiently large compressive strains. The difference in the predictions  can again be attributed to the  aforementioned limitations of orbital-free DFT.

In summary, the results demonstrate that compressive hydrostatic strains are a key driving force for vacancy clustering in aluminum. In particular, compressive hydrostatic strains trigger atomic collapse for vacancy clusters on the (111) plane.  Notably, the heptavacancy cluster on the (111) plane collapses to form a prismatic dislocation loop  for compressive strains exceeding 5\%. The formation of such loops on the (111) plane has also been observed experimentally~\cite{PureAlExpt, AlMgAlloyExpt}, supporting the physical interpretation of the predicted collapse mechanism. Indeed, kinetic effects such as vacancy diffusion rates, cluster growth dynamics, and strain evolution under realistic loading conditions have not been studied. In addition, the influence of temperature and interactions with other defects  have been neglected. Despite these limitations, the results indicate that compressive hydrostatic strains can significantly influence the spatial distribution and stability of vacancy clusters. In situations where compressive strains result from residual stress fields, thermal mismatch, impact, or irradiation, this mechanism could promote localized defect aggregation and, in some cases, prismatic loop formation. Although the loops observed here are extremely small, their formation mechanisms are relevant at much larger length scales, as the accumulation, interaction, and growth of such loops can generate extended dislocation structures that modify bulk mechanical properties, including strength, ductility, and damage tolerance.

%%%%%%%%%%%%%%%%%%%%%%%%%%%%%%%%%%%%%%%%%%%%%%%%%%%%%%%%%%%%%%%%%%%%%%%%
%%%%%%%%%%%%%%%%%%%%%%%%%%%%%%%%%%%%%%%%%%%%%%%%%%%%%%%%%%%%%%%%%%%%%%%%
%%%%%%%%%%%%%%%%%%%%%%%%%%%%%%%%%%%%%%%%%%%%%%%%%%%%%%%%%%%%%%%%%%%%%%%%

\section{Concluding Remarks\label{Sec:Conclusions}}
In this work, we have studied strain-driven vacancy clustering in aluminum from first principles. Specifically, we have performed Kohn-Sham DFT calculations to examine the influence of hydrostatic strains on  clustering in tri-, quad-, and heptavacancies. We found that compressive hydrostatic strains are a key driving force for vacancy aggregation, particularly for collapse of  clusters on the (111) plane, consistent with prior experimental observations of vacancy clusters on this plane. In particular, we found that the heptavacancy on the (111) plane collapses to form a prismatic dislocation loop  under hydrostatic compressive strains exceeding 5\%, emphasizing the critical role of such strains in the nucleation of prismatic dislocation loops in aluminum.

The influence of hydrostatic strains on the kinetics of vacancy aggregation, as revealed by migration energy calculations, presents itself as an interesting direction for future research. The construction of a full-scale kinetic model from thermodynamic data by mapping the various pathways to specific vacancy configurations presents itself as another promising area for future research.

%%%%%%%%%%%%%%%%%%%%%%%%%%%%%%%%%%%%%%%%%%%%%%%%%%%%%%%%%%%%%%%%%%%%%%%%
%%%%%%%%%%%%%%%%%%%%%%%%%%%%%%%%%%%%%%%%%%%%%%%%%%%%%%%%%%%%%%%%%%%%%%%%
%%%%%%%%%%%%%%%%%%%%%%%%%%%%%%%%%%%%%%%%%%%%%%%%%%%%%%%%%%%%%%%%%%%%%%%%

\begin{acknowledgments}
The authors gratefully acknowledge the support of the U.S. Department of Energy, Office of Science under grant DE-SC0023445. This work was performed in part under the auspices of the U.S. DOE by Lawrence Livermore National Laboratory (LLNL) under Contract DE-AC52-07NA27344. This research was also supported by the supercomputing infrastructure provided by Partnership for an Advanced Computing Environment (PACE) through its Hive (U.S. National Science Foundation through grant MRI1828187) and Phoenix clusters at Georgia Institute of Technology, Atlanta, Georgia. Additional computational resources were provided under the Multiprogrammatic and Institutional Computing program at LLNL. The authors acknowledge the valuable comments of the anonymous referees. P.S. also acknowledges insightful discussions with Vikram Gavini.
\end{acknowledgments}
%%%%%%%%%%%%%%%%%%%%%%%%%%%%%%%%%%%%%%%%%%%%%%%%%%

\section*{Data availability statements}
The data that support the findings of this article are openly available \cite{paper_data}.

%%%%%%%%%%%%%%%%%%%%%%%%%%%%%%%%%%%%%%%%%%%%%%%%%%
\appendix
\section{Vacancy formation energy \& enthalpy \label{App:VFE}} 

We now present results for the vacancy formation energy  and formation enthalpy, calculated using Eqs.~\ref{Eq:EvfV} and \ref{Eq:Hvf}, respectively. The formation energies  are computed for all the vacancy configurations, whereas the formation enthalpies are computed for select cases of interest:  monovacancy, \texttt{NN} divacancy, \texttt{A} trivacancy, and \texttt{A1} quadvacancy.  For the formation energy, to reuse data for the defect systems from the binding energy calculations, volume scaling is instead applied to the bulk crystal for the strained configurations. As a result, the actual strain values vary slightly depending on the vacancy type, but for convenience, the results are discussed using nominal strain labels: 10\% compressive (9.90\% for mono-, 9.79\% for di-, 9.69\% for tri-, and 9.58\% for quadvacancy), 5\% compressive (4.89\% for mono-, 4.78\% for di-, 4.67\% for tri-, and 4.56\% for quadvacancy), 5\% tensile (5.12\% for mono-, 5.24\% for di-, 5.37\% for tri-, and 5.49\% for quadvacancy), and 10\% tensile (10.13\% for mono-, 10.26\% for di-, 10.38\% for tri-, and 10.51\% for quadvacancy). These nominal strain labels are used for convenience in discussing the results, where the emphasis is on capturing the variation with strain rather than the precise strain values. The results presented here focus on the unstrained and compressive strain cases.  The detailed data, including results for tensile strains, are provided in the Supplementary Material \cite{Supplemental}.

\begin{figure}[htbp]
\includegraphics[width=0.44\textwidth, keepaspectratio]{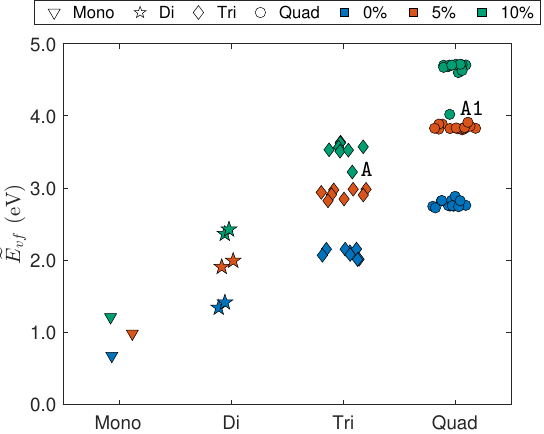}
\caption{Formation energy for the various vacancy configurations in the unstrained, 5\% compressive, and 10\% compressive strain cases. See text for the exact strain values. \label{Fig:VFE}}
\end{figure}

In Fig.~\ref{Fig:VFE}, we present the formation energy for the mono-, di-, tri-, and quadvacancies. We observe that the formation energy is positive in all cases and increases with increasing compressive strain, indicating that vacancy formation in the bulk becomes less favorable under such conditions. In contrast, the formation energy decreases with tensile strain, suggesting that vacancy formation becomes more favorable. These trends are in agreement with those predicted by orbital-free DFT  for mono-\cite{VikramPRL_Strain, CarterTriPCCP, ghosh2019electronic}, di- \cite{VikramPRL_Strain, CarterTriPCCP}, and trivacancies \cite{CarterTriPCCP}.  We also observe that the \texttt{A} trivacancy and \texttt{A1} quadvacancy are outliers under 10\% compressive strain, exhibiting significantly lower formation energies compared to the other tri- and quadvacancies, respectively. We have verified that this remains the case upon changing the exchange-correlation functional to LDA or upon using the  \texttt{PseudoDOJO} pseudopotential.

The vacancy formation enthalpy is in very good agreement with the formation energy (Supplementary Material \cite{Supplemental}), indicating that the higher order effects do not play a significant role in the formation enthalpy for the cases considered.  As a result, the sign and variation of the formation enthalpy with strain follows the same trend as the formation energy. Indeed, the \texttt{A} trivacancy and \texttt{A1} quadvacancy  are also outliers in the formation enthalpy values for 10\% compressive strain, further confirming the physical significance of the results. 

The unstrained monovacancy formation energy/enthalpy  of $0.68$~eV is in excellent agreement with the experimental value of $0.67 \pm 0.03$ eV \cite{ullmaier1991atomic}. We find that changing the exchange-correlation functional to LDA yields a formation energy of $0.72$ eV, while changing the pseudopotential to the \texttt{PseudoDOJO} version yields a value of $0.68$ eV, confirming that the results are essentially independent of the two main approximations used in Kohn-Sham DFT. The results are also in good agreement with  previous Kohn-Sham DFT calculations in literature \cite{CarlingAlVac, Carling2, UesugiAlVac, CarterTriPCCP, VacCluster8DFT}, where values ranging from $0.53$ to $0.86$ eV have been reported. The unstrained divacancy formation energies of $1.42$ and $1.34$ eV for the \texttt{N} and \texttt{NN} configurations, respectively, are in good agreement with previous DFT studies \cite{VacCluster8DFT}, which report values of $1.29$ eV for \texttt{N} and $1.24$ eV for \texttt{NN}. The formation enthalpy for \texttt{NN} is $1.37$\,eV, in good agreement with a previous Kohn-Sham DFT study~\cite{UesugiAlVac}, which reports a value of $1.40$\,eV. The unstrained formation energies of the \texttt{A} trivacancy and \texttt{A1} quadvacancy, computed as $2.12$ and $2.83$ eV, respectively, are in good agreement with previous Kohn-Sham DFT values of $2.24$ and $2.26$ eV \cite{VacCluster8DFT}. The differences between the values reported in previous Kohn-Sham calculations and those obtained in the present work can be attributed, in part, to finite-size effects in the earlier studies. The unstrained trivacancy vacancy formation energy values are also in reasonable agreement with previous orbital-free DFT calculations \cite{CarterTriPCCP}, with the maximum difference being $0.70$ eV for the \texttt{H} trivacancy. These differences, which become even more pronounced under applied strain, are likely a consequence of the aforementioned limitations of orbital-free DFT. 

In the case of the heptavacancy on the (111) plane, the vacancy formation energy is computed to be $2.937$, $5.899$, and $6.382$~eV for the unstrained, 5\% compressive, and 10\% compressive strain cases, respectively. The trend is consistent with that observed for the other vacancy clusters.

%%%%%%%%%%%%%%%%%%%%%%%%%%%%%%%%%%%%%%%%%%%%%%%%%%
\bibliography{Bibliography}% Produces the bibliography via BibTeX.
%%%%%%%%%%%%%%%%%%%%%%%%%%%%%%%%%%%%%%%%%%%%%%%%%%

\end{document}